\documentclass[preprint,aps]{revtex4}
\begin{document}
\draft
\preprint{}
\title{Electromagnetism with Magnetic Charge and Two Photons}
\author{D. Singleton}
\address{Department of Physics, University of Virginia, 
Charlottesville, VA 22901}
\date{\today}

\begin{abstract}
The Dirac approach to include magnetic charge in Maxwell's 
equations places the magnetic charge
at the end of a string on which the the fields of the
theory develop a singularity. In this paper an alternative formulation
of classical electromagnetism with magnetic and electric charge is
given by introducing a second pseudo four-vector potential, $C_{\mu}$,
in addition to the usual four-vector potential, $A_{\mu}$. This avoids 
the use of singular, non-local variables ({\it i.e.} Dirac strings) in 
electrodynamics with magnetic charge, and it makes the treatment of 
electric and magentic charge more symmetric, since both charges are 
now gauge charges.
\end{abstract}
\pacs{}
\maketitle
\newpage
\narrowtext
\section{Introduction}
Looking at the three vector form of Maxwell's equations,
one immediately notices that they become more symmetric 
if magnetic, as well as electric, charges are included. This
symmetry between electric and magnetic charges
is called the dual symmetry of electromagnetism, and
it allows one to ``rotate'' electric and magnetic quantities
({\it i.e.} fields and charges) into one another in a manner
analogous to a planar rotation \cite{jackson}. This symmetry 
does not carry over so well at the level of the gauge potential,
$A_{\mu} = ( \phi _e , {\bf A})$, since the electric and magnetic
fields look quite different in terms of these fields. Dirac 
\cite{dirac} was able to include magnetic charge in Maxwell's
equations, while keeping the usual definitions of the {\bf E} and
{\bf B} fields in terms of the gauge potentials, at the expense
of having a string singularity in the vector potential, {\bf A}.
In order to keep this string from having any
physical effect it is necessary to impose a
quantization condition on the magnetic and electric charges.
Since Dirac's initial work there has been much theoretical 
work devoted to magnetic monopoles, such as the
Wu-Yang fiber bundle formulation of magnetic charge \cite{wu}
and the 't Hooft-Polyakov monopole \cite{thooft} where a magnetic
charge arises from a non-Abelian gauge theory. In this paper we
present an alternative formulation which does not require a
Dirac string, but rather involves only local, non-singular
variables. This alternative formulation of Maxwell's equations
with magnetic charge requires the introduction of a second 
four-vector potential, $C_{\mu} = (\phi _m , {\bf C})$, 
\cite{cabibbo} in addition to 
$A_{\mu}$. The usual definitions of the {\bf E} and
{\bf B} fields are expanded so that they contain terms involving
this new potential. When these new definitions are inserted
into Maxwell's equations one finds that the equations separate
into wave-type equations for $A_{\mu}$ and $C_{\mu}$ whose
sources are electric and magentic charges respectively. This
implies that when the theory is quantized there will be
two photons -- one associated with
$A_{\mu}$ and the other with $C_{\mu}$ \cite{hagen}. However
the two photons must have opposite behaviour under the parity
or spatial inversion transformation. The advantages of 
this approach to magnetic charge are the avoidance of the string
singularity in the fields, and the expansion of the
dual symmetry to the level of the gauge potentials. Two photons
arise in this theory, since electric and magnetic charges are
now both being treated as gauge charges, which means they must
each have an associated gauge boson. The obvious objection to this
approach is that there is apparently only one massless photon in
nature. Attempts have been made \cite{zwang} to get around this
objection by introducing extra conditions on the two gauge potentials
such that the number of degrees of freedom is reduced to that
of only one photon.
In this paper we take the approach that the second gauge 
potential does represent a real gauge boson, and show what 
classical electrodynamics would look like in this hypothetical,
two-photon world by examining how 
various equations and quantities, familiar
from electromagnetism with only electric charge, change
({\it e.g.} Maxwell's equations, the Lorentz force equation,
the energy-momentum tensor). We will present most of our results
both in three-vector and in covariant four-vector notation.
The Hamiltonian formulation of the two-potential theory as well
as a discussion of some of the quantum mechanical aspects of the
problem has been given by Barker and Graziani \cite{bar}.
In the last section we will look at
some of the peculiar features of magnetic charge theory which
persist even in the two-photon model, and we will point out 
differences and similarities with the Dirac string approach. 
The fact that this second, parity odd photon is not seen 
should not immediately disqualify this theory from futher 
consideration, since one could easily use spontaneous symmetry 
breaking in the form of the Higgs mechanism \cite{higgs} to 
make this second photon massive \cite{sing}, thus hiding it,
and the gauge symmetry associated with it until a certain energy
scale had been reached (much in the same way that the $W$ and $Z$
bosons were not observed directly until a certain accelerator
energy had been reached).
In this article we are not concerned
with creating a realistic model, we simply wish to present
an alternative, and largely unknown, formulation of
electrodynamics with electric and magnetic charge, whose basic 
concepts can be grasped with an undergraduate knowledge of
electromagnetism. The results derived here are mostly a
collection of previous results. However our 
emphasis that the second potential should be treated as a real, 
second photon is unique. Throughout this paper we set $c= 1$.

\section{The Dual Photon}

The generalized Maxwell equations in the presence of electric 
and magnetic, charges and currents are \cite{jackson}
\begin{eqnarray}
\nabla \cdot {\bf E} = \rho _e  \; \; \; \; \;  \nabla \times {\bf B} =
{\partial {\bf E} \over \partial t} + {\bf J} _e \nonumber \\
\nabla \cdot {\bf B} = \rho _m  \; \; \; \; \;  - \nabla \times {\bf E} =
{\partial {\bf B} \over \partial t} + {\bf J} _m
\label{maxgen}
\end{eqnarray}
These equations possess a dual symmetry between electric and
magnetic quantities which can be written as
\begin{eqnarray}
\label{dualeb}
{\bf E} \rightarrow {\bf E} cos \theta + {\bf B} sin \theta 
\; \; \; \; \; \; \;
{\bf B} \rightarrow -{\bf E} sin \theta + {\bf B} cos \theta
\end{eqnarray}
and
\begin{eqnarray}
\label{dualeg}
\rho _e \rightarrow \rho _e cos \theta + \rho _m sin \theta
\; \; \; \; \rho_m \rightarrow -\rho _e sin \theta +
\rho _m cos \theta \nonumber \\
{\bf J}_e \rightarrow {\bf J} _e cos \theta + {\bf J}_m
sin \theta \; \; \; \; {\bf J} _m \rightarrow -{\bf J}_e
sin \theta + {\bf J}_m cos \theta
\end{eqnarray}
The {\bf E} and {\bf B}-field vectors are distinguished from one
another under the spatial inversion or
parity transformation (${\bf r} \rightarrow
-{\bf r}$) -- {\bf E} being a vector (${\bf E} \rightarrow
-{\bf E}$) and {\bf B} being a pseudovector (${\bf B} \rightarrow
{\bf B}$). These definitions and Maxwell's equations
then imply that ${\bf J}_e$ $({\bf J}_m)$ is a vector (pseudovector)
under parity, while $\rho _e$ $(\rho _m)$ is a scalar (pseudosclar)
under parity ({\it i.e.} $\rho _e \rightarrow \rho _e$ and
$\rho _m \rightarrow - \rho _m$ under  the transformation ${\bf r}
\rightarrow -{\bf r}$). In order that the parity properties of
all the quantities in the Maxwell equations 
remain unchanged under the dual rotations of Eqs. (\ref{dualeb})
(\ref{dualeg}), one must require that $\theta$ be a pseudoscalar.
Then $cos  \theta$ is a scalar since it contains only even powers
of $\theta$, while $sin \theta$ is a pseudoscalar since it contains
only odd powers.  
Our chief requirement will be that this formulation of magnetic
charge should obey this dual symmetry at every level.
Introducing two, four-vector potentials - $A ^{\mu} =
(\phi _e , {\bf A})$ and $C ^{\mu} = (\phi _m , {\bf C})$ - 
the ${\bf E}$ and ${\bf B}$ fields can be written as
\begin{eqnarray}
\label{ebpoten}
{\bf E} = - \nabla \phi _e - {\partial {\bf A} \over
\partial t} - \nabla \times {\bf C} \nonumber \\
{\bf B} = -\nabla \phi _m - {\partial {\bf C} \over
\partial t} + \nabla \times {\bf A}
\end{eqnarray}
In electrodynamics with only electric charge, {\bf E} consists
of only the first two terms, while {\bf B} consists of only the
last term. Since we still want {\bf E} and {\bf B} to be a vector
and a pseudovector under these expanded definitions we need to
require that $\phi _e$ and ${\bf A}$ act as scalars and
vectors, and $\phi _m$ and ${\bf C}$ act as pseudoscalars and
pseudovectors under spatial inversion. Inserting
these new definitions for the electromagnetic fields into Maxwell's
equations they become (after the use of a few standard vector
identities such as $\nabla \cdot [\nabla \times {\bf V}] = 0 ,
\nabla \times [\nabla \phi ] = 0$ and $\nabla \times [\nabla \times
{\bf V} ] = \nabla [\nabla \cdot {\bf V}] - \nabla ^2 {\bf V}$)
\begin{eqnarray}
\label{max1}
&&\nabla ^2 \phi _e + {\partial \over \partial t}
(\nabla \cdot {\bf A} ) = - \rho _e \; \; \; \; \; \; \; \; \; \;
\; \; \; \; \; \; \; \; \; \; \;  \; \; \; \; \;
\; \; \; \nabla ^2 \phi _m + {\partial \over \partial t} (\nabla
\cdot {\bf C} ) = - \rho _m \nonumber \\
&&\nabla ^2 {\bf A} - {\partial ^2 {\bf A} \over
\partial t^2} - \nabla \left( \nabla \cdot {\bf A} +
{\partial \phi _e \over \partial t} \right) = - {\bf J}_e
\; \; \; \; \; \;
\nabla ^2 {\bf C} - {\partial ^2 {\bf C} \over
\partial t^2} - \nabla \left( \nabla \cdot {\bf C} + 
{\partial \phi _m \over \partial t} \right) = - {\bf J}_m
\end{eqnarray}
In order to simplfy this form of Maxwell's equations further we
can use the gauge freedom possessed by the potentials,
and impose the Lorentz gauge condition on them.
\begin{eqnarray}
{\partial \phi _e \over \partial t} + \nabla \cdot
{\bf A} = 0 \nonumber \\
{\partial \phi _m \over \partial t} + \nabla \cdot
{\bf C} = 0
\end{eqnarray}
Using these gauge conditions in Eq. (\ref{max1}) we arrive at the
wave form of Maxwell's equations with electric and magnetic charge
\begin{eqnarray}
\label{max2}
\nabla ^2 \phi _e - {\partial ^2 \phi _e 
\over \partial t^2} = - \rho _e \; \; \; \; \nabla ^2
\phi _m - {\partial ^2 \phi _m \over \partial t^2}
= - \rho _m \nonumber \\
\nabla ^2 {\bf A} - {\partial ^2 {\bf A} \over
\partial t^2} = - {\bf J}_e \; \; \; \;
\nabla ^2 {\bf C} - {\partial ^2 {\bf C} \over
\partial t^2} = - {\bf J}_m
\end{eqnarray}
These equations are the inhomogeneous wave equation form of
Maxwell's equations. Notice that the parity of the quantities
on the left in Eq. (\ref{max2}) agrees with the parity of
the quantities on the right. 
In this approach to magnetic charge one has two four-vector
gauge potentials and therefore when these fields are quantized
({\it i.e.} second quantization) there are two distinct
photons. The difference between the two photons 
represented by $A_{\mu}$ and $C_{\mu}$, is that they transform 
differently under parity -- $A_{\mu}$ transforms as a 
normal four-vector while $C_{\mu}$ transforms
as a pseudo four-vector. The photon associated with $C_{\mu}$ 
will be called the dual or magnetic photon. 

All of the results up to this point can be rewritten in four-vector
notation. First we define two field strength tensors and their 
duals
\begin{equation}
F ^{\mu \nu} = \partial ^{\mu} A ^{\nu} - 
\partial ^{\nu} A ^{\mu} \; \; \; \; \;
G ^{\mu \nu} = \partial ^{\mu} C ^{\nu} -
\partial ^{\nu} C ^{\mu}
\end{equation}
\begin{equation}
{\cal F} ^{\mu \nu} = {1 \over 2} \epsilon ^{\mu \nu
\alpha \beta} F_{\alpha \beta} \; \; \; \; \;
{\cal G} ^{\mu \nu} = {1 \over 2} \epsilon ^{\mu \nu
\alpha \beta} G_{\alpha \beta}
\label{fst}
\end{equation}
where $\epsilon ^{\mu \nu \alpha \beta}$ is the 
rank four Levi-Civita tensor with $\epsilon ^{0123} 
= +1$, and having total antisymmetry in its indices.
We can write the {\bf E} and {\bf B} fields, of 
Eq. (\ref{ebpoten}), in terms of these field strength tensors
\begin{equation}
\label{ebcov}
E_i = F^{i0} - {\cal G}^{i0} \; \; \; \; \;
B_i = G^{i0} + {\cal F}^{i0}
\end{equation}
The dual symmetry can now be carried over to the potentials
\begin{eqnarray}
\label{dualac}
A_{\mu} \rightarrow A_{\mu} cos \theta + C_{\mu} sin \theta
\nonumber \\
C_{\mu} \rightarrow -A_{\mu} sin \theta + C_{\mu} cos \theta
\end{eqnarray}
The dual symmetry for the charges and currents can also be 
written in four-vector language as
\begin{eqnarray}
\label{dualcc}
J_e ^{\mu} \rightarrow J_e ^{\mu} cos \theta + J_m ^{\mu}
sin \theta \nonumber \\
J_m ^{\mu} \rightarrow -J_e ^{\mu} sin \theta + J_m ^{\mu}
cos \theta
\end{eqnarray}
where $J_e ^{\mu} = (\rho _e , \bf {J} _e)$ and $J_m ^{\mu} =
(\rho _m , \bf {J} _m)$ are the electric and magnetic four-currents
respectively.
Finally, the wave equation form of Maxwell's equations that
result from substituting Eq.(\ref{ebpoten}) into Eq. (\ref{maxgen})
are
\begin{eqnarray}
\label{maxcov}
\partial _{\mu} F^{\mu \nu} = \partial _{\mu} \partial ^{\mu}
A^{\nu} = J_e ^{\nu} \nonumber \\
\partial _{\mu} G^{\mu \nu} = \partial _{\mu} \partial ^{\mu}
C^{\nu} = J_m ^{\nu}
\end{eqnarray}
where the Lorentz gauge condition ($\partial _{\mu} A^{\mu} 
= \partial _{\mu} C^{\mu} =$ 0) has been taken for both potentials.
Since $C ^{\mu}$ is a pseudo four-vector, Eq. 
(\ref{maxcov}) implies that $J_m ^{\mu}$ must be a pseudo 
four-current. One can immediately write down a Lagrange density
which yields Eq. (\ref{maxcov})
\begin{equation}
\label{lagmax}
{\cal L} _M = -{1 \over 4} F_{\mu \nu} F^{\mu \nu} - J_e ^{\mu}
A_{\mu} -{1 \over 4} G_{\mu \nu} G^{\mu \nu} - J_m ^{\mu} C_{\mu}
\end{equation}
One could add terms like $F_{\mu \nu} {\cal G}^{\mu \nu}$,
${\cal F}_{\mu \nu} G^{\mu \nu}$, ${\cal F}_{\mu \nu} F^{\mu \nu}$,  
or ${\cal G}_{\mu \nu} G^{\mu \nu}$ to this Lagrangian and still 
obtain the Maxwell's equations of Eq. (\ref{maxcov}) since
these extra terms are total divergences due to the antisymmetry
of $\epsilon _{\mu \nu \alpha \beta}$. In the following 
section we will discuss the Lorentz force equations 
for magnetic charge. We will also construct 
the energy-momentum tensor in terms of
{\bf E} and {\bf B} fields, and in terms of
$A_{\mu}$ and $C_{\mu}$. 

It is interesting to observe that one can also think of formulating
a theory of magnetic charge with only an additional scalar potential
({\it e.g.} $\phi _m$) in addition to the standard four-vector 
potential $A_{\mu} = (\phi _e , {\bf A})$. The normal four-vector
potential, $A_{\mu} = ( \phi _e ,{\bf A})$, takes care of the {\bf E}
field (both the curl-free and divergence-free parts) as well as the 
divergence-free part of the {\bf B} field. The curl-free part of the
{\bf B} field can then be accounted for by adding only an extra
scalar, $\phi _m$, to the theory. This is in fact the basic idea
behind the 't Hooft-Polyakov \cite{thooft} monopole, where the
extra scalar potential is the Higgs field. Looking at 't Hooft's
generalized definition of the field strength tensor one sees that
the magnetic charge does indeed come from the scalar Higgs field 
rather than from the gauge fields of the theory \cite{arafune}. However 
for the 't Hooft-Polyakov monopole one finds that the magnetic charge
is not a gauge charge but a topological charge, which is not explicitly
associated with any symmetry. The present formulation of magnetic
charge is not as economical as that of 't Hooft and Polyakov, but
it is more symmetric since both electric and magnetic charges are
treated as gauge charges.

\section{The Lorentz Force Equation and Energy-Momentum Tensor}

The equation for the rate of change of mechanical energy
for an electric charge $e$ moving with velocity ${\bf v}_e$
in external {\bf E} and {\bf B} fields is
\begin{equation}
\label{eengr}
{d E_e \over dt} = e {\bf v}_e \cdot {\bf E}
\end{equation}
The Lorentz force equation for this particle is then
\begin{equation}
\label{elor}
{d {\bf p}_e \over dt} = e({\bf E} + {\bf v}_e \times {\bf B})
\end{equation}
These two equations can be combined into one manifestly covariant
form as \cite{rohr}
\begin{equation}
\label{elor1}
{d p ^{\mu}_e \over d \tau} = m {d U^{\mu} \over d \tau} =
e( F^{\mu \nu} - {\cal G}^{\mu \nu}) U_{\nu}
\end{equation}
where $U^{\mu} = d x^{\mu} / d \tau$ 
is the four-velocity of the particle with
charge $e$ and mass $m$. $\tau$ is the proper time of the
particle. Usually the covariant expression of the Lorentz force
equation involves only the $F^{\mu \nu}$ term, but because of
the expanded definitions of {\bf E} and {\bf B} one has the
second term. In electrodynamics with only electric charge it
is possible to derive the covariant form of the Lorentz force
from a Lagrangian \cite{jackson}. In the two potential
formulation  it has been proven \cite{rohr} that it is impossible,
from a single Lagrangian, to derive both the equations for the
fields and for the particles. In this paper we will get around
this problem by requiring that the generalized Lorentz equations
should satisfy the dual symmetry of Eq. (\ref{dualeb}) or 
(\ref{dualac}). Using this requirement and Eqs. 
(\ref{eengr}), (\ref{elor}) and
(\ref{elor1}) one arrives at the Lorentz equation for a
magnetically charged particle. Looking at the dual
rotation where $\theta = 90 ^o$ so that ${\bf E} \rightarrow
{\bf B}$, ${\bf B} \rightarrow - {\bf E}$, and $J_e ^{\mu}
\rightarrow J_m ^{\mu}$ ({\it i.e.} $e \rightarrow g$ and
$e {\bf v}_e \rightarrow g {\bf v}_m$) it is found that the
rate of energy change equation and the Lorentz force equation
become
\begin{equation}
\label{bengr}
{d E_m \over d t} = g {\bf v}_m \cdot {\bf B}
\end{equation}
and
\begin{equation}
\label{blor}
{d {\bf p}_m \over dt} = g({\bf B} - {\bf v}_m \times {\bf E})
\end{equation}
These can be written in covariant form as \cite{rohr}
\begin{equation}
\label{blor1}
{d p_m ^{\mu} \over d \tau} = g (G ^{\mu \nu} +
{\cal F}^{\mu \nu}) U _{\nu}
\end{equation}
For a particle carrying both types of charges ({\it i.e.} a dyon)
the equation of motion is given by the sum of Eqs. (\ref{elor})
and (\ref{blor}), or covariantly by Eqs. (\ref{elor1}) and
(\ref{blor1}).

From these Lorentz equations one can
arrive at the proper energy-momentum tensor. The time 
components of Eqs. (\ref{elor1}) and (\ref{blor1}) give 
a statement of energy conservation for a system
of particles and fields. In three vector form this yields
\begin{equation}
\label{enpoty}
{dE \over dt} = \int ({\bf J}_e \cdot {\bf E} +
{\bf J}_m \cdot {\bf B}) d^3 x
\end{equation}
Since we are now considering a system of an arbitrary
number of particles we have made the replacements
$e {\bf v}_e \rightarrow \int {\bf J}_e d^3 x$ and
$g {\bf v}_m \rightarrow \int {\bf J}_m d^3 x$.
$E$ is the mechanical energy of the particles, and the
right hand side of the equation is the rate at which external
${\bf E}$ and ${\bf B}$ fields do work on the electric and
magnetic charges. Using Maxwell's equations (written in
terms of the ${\bf E}$ and ${\bf B}$ fields) to replace ${\bf J} _e$
and ${\bf J} _m$, and applying some standard vector identities
it is possible to rewrite Eq. (\ref{enpoty}) as
\begin{equation}
\label{enpoty1}
{dE \over dt} = - {\partial \over \partial t} \int
{1 \over 2} ( {\bf E} ^2 + {\bf B}^2 ) d^3 x - \int
{\bf \nabla \cdot (E \times B)} d^3 x
\end{equation}
By making the following definition
\begin{equation}
\label{set}
T^{00} = {1 \over 2} ({\bf E} ^2 + {\bf B} ^2)
\; \; \; \; 
T^{i0} = ({\bf E} \times {\bf B}) _i
\end{equation}
one can rewrite Eq. (\ref{enpoty1}) in the suggestive form
\begin{equation}
\label{enpoty2}
{dE \over dt} = - \int  {\partial T^{00} \over \partial t}
d^3 x  - \int \partial _i T^{i0}  d^3 x
\end{equation}
This form of the energy conservation equation shows the connection
of the particles mechanical energy to the components of
the energy-momentum tensor for the fields. By moving $T^{00}$
to the left hand side of the equation it can be interpreted
as the energy density of the ${\bf E}$ and ${\bf B}$ fields.
The quantity $T^{i0}$
appears as a three divergence in Eq. (\ref{enpoty2}), which
can be converted to a surface integral, thus allowing 
$T^{i0}$ to be interpreted as an energy flux (the Poynting vector). 
These are exactly the same expressions for the field energy
density and energy flux as in the theory with only electric charge.
The difference lies in the definitions of the {\bf E} and {\bf B}
fields which have the expanded forms of Eq. (\ref{ebpoten}). This
difference will become apparent when we write down the covariant
form of the energy-momentum tensor in terms of the gauge potentials.    
Now by looking at the spatial components of Eqs. (\ref{elor1})
and (\ref{blor1}) 
it is possible to obtain a momentum conservation equation for a
system of particles interacting with external fields. Written in
three-vector form this becomes
\begin{equation}
{d{\bf P} \over dt} = \int (\rho _e {\bf E} + {\bf J}_e 
\times {\bf B} + \rho _m {\bf B} - {\bf J}_m \times {\bf E})
d^3 x
\end{equation}
where ${\bf P}$ is the mechanical momentum of the particles.
Again by using Maxwell's equations it is possible to replace
the charge and current densities ($\rho _e , \rho _m , {\bf J} _e ,
{\bf J} _m$) by derivatives of the fields to get
\begin{eqnarray}
{d {\bf P} \over dt} &=& - {d \over dt} \int ({\bf E} \times
{\bf B}) d^3 x  \nonumber \\
&+& \int [ {\bf E}(\nabla \cdot {\bf E}) -
{\bf E} \times (\nabla \times {\bf E}) + {\bf B}
(\nabla \cdot {\bf B}) - {\bf B} \times (\nabla \times {\bf B}) ]
d^3 x
\end{eqnarray}
By making the following definitions 
\begin{eqnarray}
\label{set1}
T^{0i} & \equiv & ({\bf E} \times {\bf B}) _i
\nonumber \\
T^{ij} & \equiv & E_i E_j + B_i B_j - {1 \over 2} \delta _{ij}
({\bf E} ^2 + {\bf B} ^2)
\end{eqnarray}
the equation for momentum conservation can be written as
\begin{equation}
\label{mpoty1}
{d P_i \over dt} + {d \over dt} 
\int T^{0i} d^3 x
= \int {\partial T_{ij} \over \partial x_j} d^3 x
\end{equation}
This shows the connection between the particles mechanical
momentum and the components of the energy-momentum tensor
for the fields.
Again the equation for momentum conservation has exactly
the same form as the theory with only electric charge. The
$T^{0i}$ term on the left hand side of the equation can
be interpreted as the momentum carried by the fields. The term on
the right hand side is a three divergence, which can be 
written as a surface intergral.
\begin{equation}
\oint T_{ij} n_j da
\end{equation}
Where $n_j$ is the $j^{th}$ component of the unit outward
normal to the surface. Thus $T_{ij} n_j$ can be 
interpreted as $i^{th}$ component of the flow per unit 
area of momentum across the surface. Eq. (\ref{mpoty1})
is the statement of momentum conservation of the system
of particles and fields. Collecting all the various
components of the energy-momentum tensor for the fields
one has
\begin{eqnarray}
\label{emt3df}
T^{00} &=& {1 \over 2} ({\bf E} ^2 + {\bf B} ^2) \; \; \; \; \;
T^{i0} = T^{0i} = ({\bf E} \times {\bf B}) _i
\nonumber \\
T^{ij} &=& E_i E_j + B_i B_j - {1 \over 2} \delta _{ij}
({\bf E} ^2 + {\bf B} ^2)
\end{eqnarray}
This is exactly the same form as the energy-momentum tensor
for electrodynamics with only electric charge. However the
covariant expression of the energy-momentum tensor, which is
written in terms of the gauge potentials, has extra cross
terms between the two potentials, $A_{\mu}$ and $C_{\mu}$.
Inserting the expanded definitions of {\bf E} and {\bf B} in
terms of $A_{\mu}$ and $C_{\mu}$ into the expressions of
Eq. (\ref{emt3df}) and piecing together the results
gives the covariant form of the energy-momentum tensor.
\begin{eqnarray}
\label{finemt}
T^{\alpha \beta} &=& F^{\alpha} _{\; \; \rho} F^{\rho \beta}
+ {1 \over 4} g^{\alpha \beta} F_{\mu \nu} F^{\mu \nu}
+ G^{\alpha} _{\; \; \rho} G^{\rho \beta} + {1 \over 4}
g^{\alpha \beta} G_{\mu \nu} G^{\mu \nu} \nonumber \\
&+& {1 \over 2} ({\cal F} ^{\alpha \rho} G_{\rho} ^{\; \; \beta}
+ {\cal F} ^{\beta \rho} G_{\rho} ^{\; \; \alpha})
- {1 \over 2} ({\cal G} ^{\alpha \rho} F_{\rho} ^{\; \; \beta}
+ {\cal G} ^{\beta \rho} F_{\rho} ^{\; \; \alpha})
\end{eqnarray}
This is the covariant expression for the energy-momentum tensor
of Ref. \cite{rohr}. It is symmetric in its indices $\alpha$ and
$\beta$ as is required if the angular momentum of the fields is
to be conserved.
The top line of the right hand side of the equation are the terms
one would expect by a simple generalization of the usual
energy-momentum tensor for electrodynamics with only electric charge.
The terms on the second line are the cross terms between $A_{\mu}$
and $C_{\mu}$ which occur when one inserts the expressions for
{\bf E} and {\bf B} from Eq. (\ref{ebpoten}) into Eq. (\ref{emt3df}).
This form of the energy-momentum tensor ensures that the magnetic
and electric charges will interact with one another according to
the Lorentz force equations of Eqs. (\ref{elor1}) and (\ref{blor1}).
These cross terms also ensure that the combination of an electric
charge and a magentic charge will carry some angular momentum in
their combined {\bf E} and {\bf B} fields \cite{saha}. One can 
reverse the arguments above and starting with the energy-momentum
tensor of equation Eq. (\ref{finemt}) arrive at the Lorentz equations.
Taking the four divergence of $T_{\mu \nu}$ and using the covariant
form of Maxwell's equations gives, after some work
\begin{equation}
\partial _{\mu} T^{\mu \nu} = -F^{\nu \rho} J^e _{\rho} - G^{\rho \nu}
J^m _{\rho} -{\cal F} ^{\nu \rho} J^m _{\rho} + {\cal G}^{\nu \rho}
J^e _{\rho}
\end{equation}
This shows that in the absence of external sources that the
energy-momentum tensor is divergenceless, and in the presence
of external sources the above equation gives the covariant form
of the generalized Lorentz force of Eqs. (\ref{elor1}) and
(\ref{blor1}) if one includes the mechanical energy-momentum 
tensor for the particles.

\section{Angular Momentum and Quantization Conditions}

In this section we examine some of the peculiar aspects of
monopole theory, which occur in this two-potential approach
as well as in other formulations. In particular 
we examine the angular momentum which is carried in the 
fields produced by two particles -- one carrying an electric
charge and the other a magnetic charge. We will briefly
discuss how the quantization condition between the two types 
of charges carries over into the two-potential theory, and
we will make some comments about symmetry breaking in magnetic
charge theories.

To study the angular momentum carried in the electromagnetic
fields of some configuration of charges one looks at the integral 
of certain components of the rank three tensor
\begin{equation}
\label{angten}
{\cal M} ^{\alpha \beta \gamma} = T^{\alpha \beta} x^{\gamma}
- T^{\alpha \gamma} x ^{\beta} 
\end{equation}
where $T^{\alpha \beta}$ is the energy momentum tensor of Eq. 
(\ref{finemt}). In particular the angular momentum is given by
\begin{equation}
\label{ang3d}
M^{ij} = \int {\cal M} ^{0ij} d^3 x = \int (T^{0i} x^j
-T^{0j} x^i ) d^3 x
\end{equation}
The configuration that we will consider is of two particles,
${\cal A}$ and ${\cal B}$, with particle 
${\cal A}$ having an electric charge $e$ (whose magnitude 
we will take to be that of an electron or proton) and particle 
${\cal B}$ having a magnetic charge $g$, which is undetermined 
at this point. In connection
with our introductory comments we will allow the magnetic photon
to have a mass $m$. Usually gauge boson are prohibited from having
a mass by gauge invariance. A massive gauge boson introduces a term
${1 \over 2} m^2 C_{\mu} C^{\mu}$ into the Lagrangian, which is
then not invariant under the gauge transformation $C_{\mu} \rightarrow
C_{\mu} - \partial _{\mu} \Lambda (x)$, where $\Lambda (x)$ is
an arbitrary function (note that the field strength tensor, 
$G_{\mu \nu}$, is invariant under this gauge transformation). 
However it is possible to use the Higgs
mechanism \cite{sing} to give the magnetic photon a mass while
still maintaining a gauge invariant theory. Here we are giving the
magnetic photon a mass by hand  with the justification that this
development can be made consistent with gauge invariance
by an application of the Higgs
mechanism. Placing the magnetic particle, ${\cal B}$, at the origin and 
the electric particle, ${\cal A}$, a distance ${\bf d}$ from ${\cal B}$ 
the four-vector potentials produced by these particles are
\begin{eqnarray}
\label{egpot}
C_{\cal B} ^{\mu} &=& \left( {g \over 4 \pi} {e^{-m r} \over r} , 0, 0, 0
\right) \nonumber \\
A_{\cal A} ^{\mu} &=& \left( {e \over 4 \pi} {1 \over r'} , 0, 0, 0 \right)
\end{eqnarray}
where $r' = \vert {\bf r} - {\bf d} \vert$. Notice that particle ${\cal A}$
produces a Coulomb potenial, while particle ${\cal B}$ produces a
Yukawa potential due to the postulated mass of of the magnetic photon.
Without loss of generality we take particle ${\cal A}$ to be located 
along the $+ z$ - axis. Then plugging the four-vector potentials of Eq.
(\ref{egpot}) into the energy-momentum tensor and finally into the
expression for the field angular momentum of Eq. (\ref{ang3d}) we find
the angular momentum of this charge configuration to be
\begin{equation}
\label{angpart}
M^{ij} = L_k = - {egd \over 16 \pi ^2} \int {e^{-mr} \over {r'} ^3}
\left( {1 \over r} + m \right) \big[n_z - n_k cos \theta \big] d^3 x
\end{equation}
where $n_k$ is the unit vector in the $k ^{th}$ direction. Evaluating
this in spherical polar coordinates and noticing that only the
$z$ component of $n_k$ survives the $\phi$ integration we arrive
at
\begin{equation}
\label{angfin}
L_z = - {egd \over 8 \pi} \int _0 ^{\infty} r^2 dr \left[ \int _{-1} ^1
d(cos \theta) {e^{-mr} \over (r^2 + d^2 - 2rd cos \theta) ^{3/2}}
\Big( m + {1 \over r} \Big) \big( 1 - cos ^2 \theta \big) n_z \right]
\end{equation}
The remaining $r$ and $\theta$ integrals can be performed  using some
standard integration techniques (see the article by Carter and
Cohen \cite{cart} for the case when $m=0$). This leads to
\begin{equation}
\label{angfin1}
L_z = - {eg \over 2 \pi m^2 d^2} \Big[ 1 - (1+md)e^{-md} \Big] n_z
\end{equation}
Thus, under the assumption that the magnetic photon has a mass
$m$, we find that the angular momentum in the charge-monopole
system depends on the magnitude of the two types of charge,
$e$ and $g$, and also on the mass, $m$, and the separation,
$d$, between the two particles. This latter feature (the dependence
of the angular momentum on the seperation between the charge
and the monopole) is a unique feature that arises because we
have assumed a mass for the magnetic photon. In the usual analysis
of the charge-monopole system \cite{cart} the angular momentum
is independent of the distance between the two particles. This
difference will yield some interesting results when we discuss the 
quantization conditions between the charges. To make the connection
with the usual angular momentum result we let $m \rightarrow 0$.
Care must be taken since $m$ occurs in the denominator of Eq.
(\ref{angfin1}). Expanding $e^{-md}$ out to second order in $md$
Eq. (\ref{angfin1}) becomes
\begin{equation}
\label{angaprox}
L_z =  \left ( -{eg \over 4 \pi} + {eg \over 4 \pi} m d 
- {\cal O} (m^2 d^2) \right) n_z
\end{equation}
On taking the limit $m \rightarrow 0$ we recover the standard result
which is independent of the distance between the two particles. We have 
calculated the angular momentum of the fields using four-vector
notation. The entire calculation can be done in three-vector form
by taking the results  for the three-vector form of $T^{0i}$ 
and plugging them into Eq.(\ref{ang3d}) to give
\begin{equation}
\label{ang3d1}
{\bf L} = \int {\bf r} \times ({\bf E} \times {\bf B}) d^3 x
\end{equation}
for the angular momentum in the {\bf E} and {\bf B} fields.
Expressions for the {\bf E} and {\bf B} fields can be obtained 
using the potentials of Eq. (\ref{egpot}).
\begin{eqnarray}
{\bf E} &=& -{e \over 4 \pi} {{\bf r'} \over {r'} ^3} \nonumber \\
{\bf B} &=& -{g \over 4 \pi} {e^{-mr} \over r^2} \left( {1 \over r}
+ m \right) {\bf r}
\end{eqnarray}
Inserting these expressions into Eq. (\ref{ang3d1}) one can go
through the same steps which were taken in the covariant notation
to rederive the result for the angular momentum in the {\bf E} and
{\bf B} fields of the configuration.

One of the most interesting results of Dirac's string monopole
theory is the quantization condition between electric and
magnetic charges
\begin{equation}
\label{dirac}
{e g \over 4 \pi} = {n \over 2}
\end{equation}
(Here we are only concerned with magnitude, as opposed to Eq.
(\ref{angaprox}) which also gives the direction of {\bf L}). $n$ is 
an integer, and we have set $\hbar = 1$ -- it would appear in the 
numerator of the right hand side of Eq. (\ref{dirac}). This condition
arises from the requirement that the wavefunction of a particle
in the presence of the Dirac string be single valued \cite{dirac}.
Thus even in a first quantized theory ({\it i.e.} where the particles
are quantized but the fields are treated classically) one has a
restriction between electric and magnetic charges. This condition
has the physical consequence that the singular string variable has no
physical effect on the particles of the theory. It makes
the string invisible to the particle. In the two-potential approach
we have replaced the non-local and singular string with a local and
non-singular gauge field. Since there is no prohibition against
having particles interact with the non-singular 
gauge field, $C_{\mu}$, one does not get a restriction 
between the electric and magnetic charges as in Eq. (\ref{dirac}). 
However when the fields, $A_{\mu}$ and $C_{\mu}$, are quantized, 
({\it i.e.} second quantization) one recovers such a condition,
although for the case when the magnetic photon is massive this condition
is considerably different from Dirac's condition of Eq. (\ref{dirac})

The basic argument, which is due to Saha and Wilson \cite{saha}, is that
when the gauge fields are treated as quantum fields, the angular
momentum carried in the field configuration of an electric
charge and a magnetic charge must be quantized in integer multiples
of $\hbar / 2 $. Taking the result of Eq. (\ref{angfin1}) in 
conjunction with this quantum restriction on the angular momentum 
yields the condition
\begin{equation} 
\label{dirac2}
{eg \over 2 \pi m^2 d^2} \Big[ 1 - (1+md)e^{-md} \Big] = {n \over 2}
\end{equation}
where $n$ is an integer. Only the magnitude of $L_z$ from Eq.
(\ref{angfin1}) is taken, and as stated previously we are setting 
$\hbar = 1$. We shall address most of our comments to the $n = 1$ 
case. In the limit $m \rightarrow 0$ we just recover the
quantization condition of Eq. (\ref{dirac}). When $m \ne 0$ we get a
new quantization condition which involves not only the magnitude
of the charges, $e$ and $g$, but the mass of the magnetic photon and
the distance between the two particles. Comparing the two conditions 
from Eq. (\ref{dirac}) and Eq. (\ref{dirac2}) one notices that the 
magnitude of the magnetic charge, $g$, must always be larger for
the latter condition. Physically this is easily understood since in
the case of the massive magnetic photon the Yukawa field of the 
magnetically charged particle falls off more rapidly than the 
equivalent Coulomb case, so in order for the configuration to
still have an angular momentum of $ 1/2 $ we need to have a larger
magnetic charge, $g$. Thus one can say that as the mass $m$ increases
the magnitude of the magnetic charge must also increase in order to
have a minimum angular momentum of $1/2$. A caveat to this
statement is that one can also change the angular momentum by changing
the distance $d$. In general one can increase the angular momentum
of the charge-monopole system by decreasing the separation between
them. Decreasing the distance between the particles allows the
electric charge to ``see'' more of the magnetic charge.
It is easy to see that the condition imposed by Eq. (\ref{dirac2})
is very complicated since it depends on three variables, $g$, $m$
and $d$ ($e$ is assumed to be fixed to the magnitude of the electron's
charge). The most unusal result however is that if one specifies
some mass for the magnetic photon and some magnitude for the magnetic
charge, then the condition of Eq. (\ref{dirac2}) restricts the separation
distances between the charge and monopole to be only certain discrete
values. This quantization of the separation distance is in a sense
a kinematic restriction, since we can think of no dynamical reason
that would force the charge-monopole system to have such discrete
separations.

To conclude this section we will give a speculative argument which
suggests that in any monopole theory the photon may acquire a mass.
Usually gauge bosons are said to be massless due the gauge invariance.
However one should add to this statement the requirement that 
the coupling constant be enough small so that perturbation theory is
valid \cite{huang}. Some good, but non-rigourous arguments have been
given by Wilson \cite{wil} and Guth \cite{guth} which suggest that
there is some critical value of electric charge below which
the photon is massless and above which the photon acquires a mass.
This critical value is not specified by their analysis,
but due to the smallness of the electric charge of all the
known particles it seems reasonable to assume that this critical
value is greater than the known electric charge magnitude. 
The success of perturbation theory for the electromagnetic 
interactions of the electron also indicates that this is a good
assumption. In quantum electrodynamics in one space and
one time dimension Schwinger \cite{schwing} has shown rigorously
that this phenomenon of the photon acquiring a mass 
does in fact occur. Although Schwinger's $1+1$ quantum electrodynamics
shows this dynamical mass generation for the photon, it is not
clear what relevance it has to theories in $3+1$ dimensions. 
In particular in Schwinger's model one finds that the critical
value of the electric charge is zero \cite{holstein} ({\it i.e.} the
photon will develop a mass if $e \ne 0$ in $1+1$ dimensions)). With 
these motivating statements let us look at the magnetic charge. Since 
we are trying to argue that the photon in any formulation of a
magnetic charge theory will become massive, we will not from the
outset assume a mass for the magnetic photon. This means that we
have the same quantization condition for both the string and the 
two-potential theories. Thus in either case the magnitude
of the magnetic charge is large ({\it i.e.} at the minimum
$g = 2 \pi / e$ which is large since $e$ is small), and
we are well out of the
regime where perturbation theory is valid. Therefore based on the
conjectures of Wilson and Guth it is not unreasonable to 
suggest that the photon may acquire a mass in the presence of
magnetic charge with such a large magnitude. This mechanism, which
could generate a mass for the photon, can be compared to the
technicolor theory of particle physics \cite{cheng}, 
which also give a mechanism for dynamically generating masses 
for the $W^{\pm}$ and $Z$ gauge bosons of the standard 
model. In technicolor theories one has techniquarks instead of
magnetic monopoles, and a super strong ``color'' force instead
of a super strong electromagnetic force.

Coupling Wilson and Guth's conjecture with the duality of
electric and magnetic charge (which implies that it should not
make a difference whether the large, non-perturbative coupling
is electric or magnetic) one is led to the conclusion that in the
presence of large magnetic charge the photon may become massive. 
This could be viewed as a loose argument against a theory of
magnetic charge with only a single photon. (The reason for saying
that this is only a loose argument, in addition to the fact that 
Wilson and Guth's conjecture has not been rigourly proven, is
that one could always claim that the mass given to the photon
is below the experimental upper limit. However, since
the upper bound on the photon's mass is very 
stringent, this is not a strong counterargument). The two photon
theory of magnetic charge however is still viable under this
conjectured mechanism for photon mass generation, since one of
the photons ({\it i.e.} the magnetic photon) can become massive
while the other photon ({\it i.e.} the electric photon) remains
massless. This is in direct analogy with the standard model
where the $Z$ boson acquires a mass while the photon remains
massless.

\section{Discussion and Conclusions}

Based on an old idea of Cabibbo and Ferrari \cite{cabibbo},
we have reviewed the covariant Lagrangian formalism of
electromagnetism with magnetic charge, which employs two 
four-vector potentials, $A_{\mu}$ and $C_{\mu}$. The Hamiltonian
formulation of this approach to magnetic charge, as well 
as some of the quantum mechanical aspects 
of this theory can be found in two articles by
Barker and Graziani \cite{bar}. This strategy has several differences
with the usual Dirac formulation of magnetic charge -- (a) It allows
one to extend the dual symmetry down to the level of the gauge 
potentials, and it treats magnetic and electric charges in the same
way ({\it i.e.} both are gauge charges). (b) It deals only with local, 
nonsingular variables. (c) Since there is no singular string in the
theory there is no quantization condition on the charges in a
first quantized theory. If the gauge fields are quantized one
recovers a quantization condition which is the same as Dirac's
condition only if the magnetic photon is massless. If the magnetic
photon is massive one obtains an unusual quantization condition which
involves not only the electric and magnetic charge magnitudes, but
also the mass of the magnetic photon as well as the distance between
the charges.
The main disadvantage of this two potential idea is that there is no
single Lagrangian from which both the field equations and the Lorentz
equations can be derived. The field equations, in terms of the
two potentials, can be derived from the Lagrangian of Eq. 
(\ref{lagmax}), which is a straight forward extension of the usual
field Lagrangian in electromagnetism with only electric charge. According
to a result by Rohrlich \cite{rohr} it is not possible to use a 
single Lagrangian to derive both the generalized Lorentz force 
equations and the field equations
when one has both electric and magnetic charges. 
Here we side step this problem by not deriving
the Lorentz equations via a Lagrangian, but rather from the requirement
that the theory respect the dual invariance of Eq. (\ref{dualeb}). 
These generalized Lorentz equations in turn allowed one to arrive at
an expression for the energy-momentum tensor.
A different approach to this problem of the Lorentz force
equations can be found in Ref. \cite{rodrig} where the Lagrangian is
modified so that it becomes possible the derive the covariant form
of the energy-momentum tensor 
directly from the Lagrangian. Then, since the
Lorentz equations imply the form of the energy-momentum tensor and
visa versa, it is possible to obtain the Lorentz equations by taking
the four-divergence of the energy-momentum tensor obtained from the
modified Lagrangian. This is similiar to the
situation in general relativity where the particle equations 
of motion follow from the field equations rather than being
distinct from them \cite{einstein}. To use this approach one can
add
\begin{equation}
{\cal L}_{extra} = -{1 \over 4} {\cal F}^{\mu \nu}(\partial _{\mu}
C_{\nu} - \partial _{\nu} C_{\mu}) +
{1 \over 4} {\cal G}^{\mu \nu}(\partial _{\mu}
A_{\nu} - \partial _{\nu} A_{\mu})
\end{equation}
to the Lagrangian of Eq. (\ref{lagmax}). As was mentioned at 
the end of section II, these added terms will not
change Maxwell's equations since they are total four-divergences.
Formally the energy-momentum tensor of Eq. (\ref{finemt}) 
is obtained by treating $A_{\mu}$, $C_{\mu}$,
${\cal F} _{\mu \nu}$ and ${\cal G} _{\mu \nu}$ as independent
variables. Then by adding four-divergences 
at the appropriate points, we
arrive at the expression from Eq. (\ref{finemt}). As already
pointed out, taking the four divergence of the energy-momentum
tensor then yields the covariant form of the Lorentz equations.
In the simple development given in this paper the general Lorentz
equations and the energy-momentum tensor are obtained by assuming
that the usual form for the Lorentz equations with only electric
charge, and then deriving the general form using the
dual symmetry. In this way the dual theory does not treat the
field equations and the particle equations equivalently since
the field equations can be obtained from a Lagrangian, while
the particle equations are gotten by requiring that the dual 
symmetry hold.

By introducing a second, pseudo four-vector potential, $C_{\mu}$,
it is possible to obtain a formulation of electrodynamics with
both electric and magnetic charges, which is an alternative
to Dirac's string approach. The basic difference between
the two-potential approach and the string approach, is that one
replaces a non-local, singular variable ({\it i.e.} the Dirac string)
with a local, non-singular variable ({\it i.e.} the pseudo
four-potential $C_{\mu}$). This two-potential approach is not new, 
but usually it is argued that the second potential does
not represent a second photon. This is done by introducing extra
conditions on the gauge potentials \cite{zwang} so that the
number of degrees of freedom reduce to those of only one photon.
Here we take the view point that this second
potential does represent a second, magnetic photon, which has
the opposite behaviour under parity as the usual photon, and we
develop what electrodynamics would look like in a hypothetical
world that did have two massless photons. To make the theory
more realistic one could easily apply the Higgs mechanism
to give the magnetic photon a mass, taking it out of the
observable particle spectrum until a certain energy scale
had been reached. Finally a speculative argument is given which
implies that the large magnitude of the magnetic charge in 
either the string approach or the two potential approach leads
to a dynamical generation of mass for the photon. In a theory 
with only one photon this is bad since we know that there is
one massless photon. In a theory of magnetic charge with two
photons however it can be arranged so that one photon becomes
massive while the other remains massless.

\section{ACKNOWLEDGEMENTS} The author would like to thank
David New and Marla Cantor for help and encouragement
during the writing of this paper. Additional thanks goes
to A. Yoshida who has discussed many of the aspects of
this paper with the author.

\end{document}